\documentclass[apj,twocolumn]{OJA}

\usepackage{natbib,graphics,epsfig}
\usepackage{newtxtext,newtxmath}

\usepackage[T1]{fontenc}
\usepackage{ae,aecompl}

\usepackage{graphicx}	
\usepackage{amsmath}	
\usepackage{graphicx}
\usepackage{epsfig}
\usepackage{natbib}
\usepackage{multirow}
\usepackage{amsmath}

\newcommand{\bal}{H$\alpha$/H$\beta$}
\newcommand{\ha}{H$\alpha$}
\newcommand{\hab}{H$\alpha$/H$\beta$}
\newcommand{\hb}{H$\beta$}
\newcommand{\hg}{H$\gamma$}
\newcommand{\hd}{H$\delta$}
\newcommand{\lya}{Ly$\alpha$}
\newcommand{\lyb}{Ly$\beta$}

\newcommand{\hi}{H\thinspace{\sc i}}
\newcommand{\hii}{H\thinspace{\sc ii}}
\newcommand{\nii}{[N\thinspace{\sc ii}]}

\newcommand{\oiii}{[O\thinspace{\sc iii}]}

\newcommand{\oii}{[O\thinspace{\sc ii}]}

\newcommand{\ariv}{[Ar\thinspace{\sc iv}]}

\newcommand{\sii}{[S\thinspace{\sc ii}]}

\newcommand{\heii}{He\thinspace{\sc ii}}
\newcommand{\hei}{He\thinspace{\sc i}}

\shorttitle{On the validity of Case~B}
\shortauthors{Scarlata et al.}

\begin{document}

\title[On the validity of Case~B]{On the universal validity of Case~B recombination theory}


\author{C. Scarlata$^{1}$}
\author{M. Hayes$^{2}$}
\author{N. Panagia$^{3,4}$}
\author{V.  Mehta$^{1}$}
\author{F. Haardt$^{5,6}$}
\author{M. Bagley$^{1}$}
\affiliation{$^{1}$Minnesota Institute for Astrophysics, University of Minnesota, Minneapolis, MN 55455, USA.}
\affiliation{$^{2}$Stockholm Observatory, Department of Astronomy, Stockholm University, 106 91 Stockholm, Sweden}
\affiliation{$^{3}$Space Telescope Science Institute, 3700 San Martin Drive, Baltimore MD 21218, USA}
\affiliation{$^{4}$Supernova Limited, OYV \#131, Northsound Rd., Virgin Gorda VG1150, Virgin Islands, UK}
\affiliation{$^{5}$DiSAT, Universit\`a dell'Insubria, via Valleggio 11, 22100 Como, Italy }
\affiliation{$^{6}$ INFN, Sezione di Milano-Bicocca, Piazza delle Scienze 3, 20123 Milano, Italy}

\email{Corresponding Author: mscarlat@umn.edu}

\begin{abstract}
In an ongoing search for low-mass extreme emission line galaxies, we identified a galaxy with a  \ha/\hb\ Balmer line ratio of $2.620 \pm 0.078$. \ha/\hb\ Balmer ratios lower than the dust-free Case~B value appear relatively frequently in extreme emission line galaxies.  These low values suggest that the Case~B assumption may not be valid in these objects. After ruling out the possibility that the low \ha/\hb\  ratio  is due to systematic errors introduced by observational effects,  we use constraints from  the total \hb\ luminosity, the \oiii/\oii\ line ratio and the Balmer line equivalent widths, to suggest that the gas is optically thick to both Balmer \ha\  and \lya\ photons, and the geometry and orientation of the scattering gas causes \ha\ photons to be preferentially removed from the line of sight with respect to higher order Balmer series photons.  Finally, we use data from the SDSS survey to show that Balmer self--absorption  may be more important than previously assumed in high excitation emission line galaxies, where \lya\  pumping of the hydrogen excited state can be effective. 
If not recognized, Balmer self-absorption could lead to  inaccurate estimates of galaxy physical properties. As an example, the effect of dust extinction could be over-estimated, for spherically symmetric scattering medium, or under-estimated, for a not spherically-symmetric distribution.  

\end{abstract}



\section{Introduction}
Over the past decade, it has become clear that the galaxies we are able to observe in, or close to, the reionization epoch are characterized by young starbursts with strong nebular emission lines,  compact sizes,  low dust content, and high ionization parameters \citep[e.g.,][]{Schaerer2009,Finkelstein2012,Dunlop2013,Capak2015,Bouwens2017,DeBarros2019,Endsley2020}. These results have now been confirmed with extensive spectroscopic campaigns conducted with the James Webb Space Telescope, which push observations of rest-frame optical lines out to redshifts $\approx 9$ \citep[e.g.,][]{Topping2024,Pirzkal2023}. 

Wide field optical surveys like the Sloan Digital Sky Survey  \citep[SDSS][]{sdss}  have shown that galaxies can be identified at low-$z$ that have properties,  comparable to these reionization era systems \citep{Schaerer2022}. These analogs are characterized by 
compact morphologies and extreme nebular spectra, with  high \oiii/\oii\ line ratios \citep[typically above 5,][]{Izotov2017} and  extreme  Balmer line equivalent widths (EWs, over several hundreds \AA\ for \ha, and \oiii).  Additionally, UV spectra of these objects show a high  incidence of \lya\ in emission (very close to 100\%),  much higher than in  normal star forming galaxies with similar stellar masses \citep[e.g.][]{Henry2015,Jaskot2019}.   These properties are indicative of low-metallicity  \citep[Z$\lesssim$10\% Z$_{\odot}$,][]{Cardamone2009,Amorin2010,Yang2017,Senchyna2017}, highly excited gas, ionized by young and hot starbursts, and with typically small amount of dust  \citep[e.g.,][]{Henry2018, Jaskot2017, Jaskot2019}.  Finally, direct measurements of their ionizing spectra (below rest--frame 912\AA) proved that  a substantial fraction of their ionizing radiation is able to escape  the interstellar medium, with escape fraction up to 70\%, \citep[][]{Izotov2016a,Izotov2016b,Schaerer2016,Verhamme2017,Izotov2018b,Flury2022a,Flury2022b}.

Many of the  physical properties derived from the emission line spectra of these galaxies are based on one fundamental assumption: that the Case~B recombination limit is valid and can be  used to interpret their spectra. Although this assumption is verified in the vast majority of star-forming galaxies, there has been some evidence that the spectra of extreme emission line galaxies (EELGs) may deviate from Case~B: 32 out of 41 of the EELGs  in \citet{Yang2017} have \ha/\hb\  ratio smaller than  (\ha/\hb)$_{\rm Case B}$. Similarly, \citet{Atek2009} find that a number of $z\sim0.3$ \lya\ emitters have comparably low Balmer line ratios.  Early results based on JWST spectra show similar trends: \citet{Pirzkal2023} find a significant number of $1\lesssim z \lesssim 3.5$ sources in the NGDEEP-NISS survey which appear to violate Case~B recombination.  This result seems to hold at even higher redshifts, where examples of \ha/\hb\ line ratios smaller than Case~B are found at $z\approx 6$ by, e.g., \citet{Cameron2023} and \citet{Topping2024}. As part of an ongoing search aimed at pushing the selection of local analogs to lower masses and lower  metallicities, described in \citet{Lin2022}, we discovered one object (hereafter, SXDF308)  that shows an emission line spectrum that could dramatically change the physical interpretation of EELG spectra. The Balmer \ha/\hb\ line ratio measured in this object suggests that the universally accepted Case~B limit is in fact not a good assumption for this galaxy.  We argue in Section~\ref{sec:discussion} that the physical conditions specific to this galaxy may in fact be more common than previously assumed.  

The implications of this finding are important: if Case~B is not a valid assumptions for some high excitation emission line galaxies, then physical properties such as dust extinction, intrinsic ionizing power, and consequently Lyman continuum escape fractions, computed from the intensity of the Balmer lines will be  incorrect. In this paper we present a thorough investigation of the emission line spectrum of SXDF308, including Balmer and Oxygen line emission, to determine the properties of the nebular gas and fully explore the reasons for the departure from the Case~B scenario. 

The paper is organized as follows: in Section~\ref{sec:data} we present the observations and  the data analysis. 
Different physical interpretations for the anomalous line ratios are addressed in Section~\ref{sec:interpretation} and discussed in Section~\ref{sec:discussion}. We present our conclusions in Section~\ref{sec:conclusions}.

\begin{figure*}
\includegraphics[width=\textwidth]{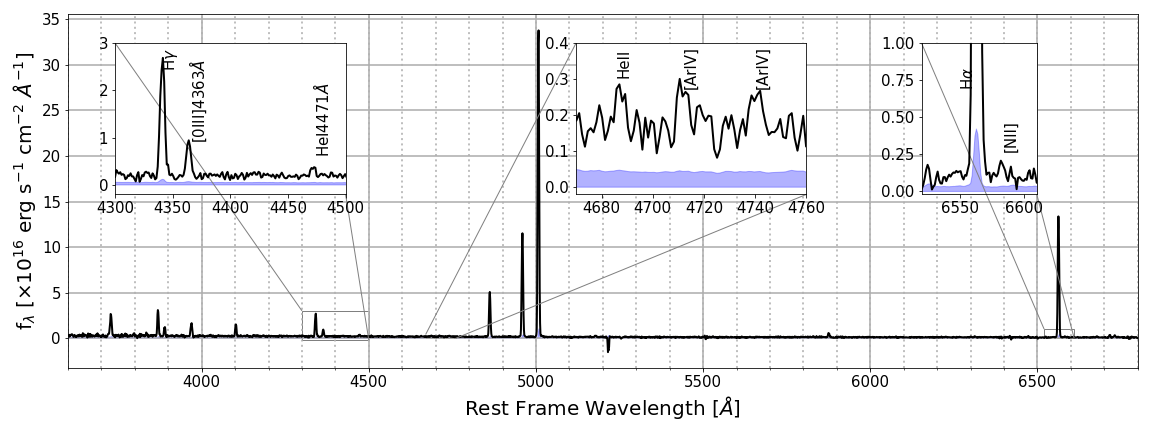}
\caption{{\it Main panel:} MMT spectrum of SXDF308. The full wavelength and flux range is shown to emphasize the line ratios between the strongest emission lines. In particular, note the large \oiii5007/\hb\ line ratio, indicating the presence of a hard ionizing radiation field. {\it Insets:}  zooms in around  faint spectral features, indicating very low abundance (e.g., low \nii/\ha\ ratio), high electron temperature (from the strong \oiii$\lambda 4363$) and hard ionizing radiation field (\heii,\ariv). The blue shaded region indicates the $1\sigma$ error on the spectrum. \label{fig:one}}
\end{figure*}

\section{Observations and data reduction}
\label{sec:data}
SXDF308 (RA=02:19:12.857, DEC=--04:30:17.62, J2000) was discovered as an extreme emission line candidate using the $u$, $g$, $r$ broad band  images covering the 2 deg$^2$ Subaru Deep Field \citep{Mehta2018}. Extreme emission line candidates are selected as objects showing a significant $g-$band flux excess, that we assume is a result of a strong emission line with observed equivalent width larger than few hundreds \AA. This filter combination identifies galaxies in the 0.05$\lesssim z \lesssim 0.1$ and 2.3$\lesssim z \lesssim 3.1$ redshift ranges, where the emission line(s) in question may be \oiii+\hb\ and \lya, respectively.  SXDF308 was spectroscopically confirmed to be a low redshift galaxy in the spectroscopic followup and has a redshift of $z=0.0695$.
 
The spectroscopic data for SXDF308 were obtained in queue mode on Sept. 26, 2017 using the Hectospec Multifiber Spectrograph at the MMT 6.5m telescope in Arizona. Hectospec uses the $f/5$ secondary mirror, which incorporates atmospheric dispersion compensation prisms to correct for atmospheric dispersion between airmass of 1 and 2. We used the 270 mm$^{-1}$ grating, blazed at 5200\AA\ to provide a spectral coverage of 3650--9000\AA\  \citep{Fabricant2005}, at a spectral resolution, for the used fiber diameter (1\farcs5), of  $\approx 6$\AA.
Second-order blocking filters are not available to use with Hectospec. However, as we discuss below,  the reddest line of interest ([S~{\sc ii}]~$\lambda 6731$~\AA) falls at 7200~\AA, where the second order contamination from blue light is negligible \citep[][]{Gezari2012}.  The total integration time was 4,500 seconds, split in 5 exposures to remove cosmic ray hits.  The seeing full-width-half-maximum during the observations was stable at 1\farcs0,  conditions were clear, and the average airmass of the observations was 1.4. A large number (103) of fibers were placed on empty sky\footnote{Empty sky regions were identified using SPLASH deep imaging} regions, resulting in an accurate determination of the sky background, with the sky fiber-to-fiber variation of  $\approx$0.2\%.

The data were reduced using the  {\fontfamily{qcr}\selectfont  HSRED 2.0}\footnote{http://archive.mmto.org/node/536} which was originally written by R. Cool. The pipeline performs bias, flat-field, illumination, and wavelength calibrations, background sky subtraction, and extracts one-dimensional spectra. The flux calibration was performed using 3 standard stars (BD+28-4211, HD-217086, HD-192281) observed on Oct. 1--3, 2017, at  airmasses between 1 and 1.4,  along with a set of 4 stars with available broad--band photometry from SPLASH-SXDF, observed together with SXDF308. We computed the wavelength dependent sensitivity function as the average of the sensitivity functions obtained from all the standard stars exposures. We verified that second-order contamination does not affect the results of the flux calibration. First, at the wavelength of interest (7020\AA\ for \ha) the contamination of the stellar continuum due to radiation at 3500\AA\ is  minimal (Fabricant, private communication). Second, as the amount of second-order contamination from blue light depends on the color of the source, we verified that the computed sensitivity curve does not depend on the color of the calibration star, demonstrating that second-order contamination is not an issue bluewards of 7200~\AA. 
At the high signal--to--noise ratio of the available data, the major uncertainty in the flux measurements comes from systematic uncertainties introduced during the flux calibration step  \citep[e.g.,][]{Berg2015}. We quantify this systematic contribution in Appendix~\ref{sec:app2}.

The extracted spectrum was corrected for Milky Way foreground extinction using the \citet{1989ApJ...345..245C} extinction law, assuming $R_V=3.1$ and an $E(B-V)=0.0182 \pm 0.0002$ computed from the \citet{Schlafly2011}  recalibration of the \citet{schlegel1998} extinction map [corresponding to $A_V =  0.0561$(mag)]. 
Figure~\ref{fig:one} shows the extracted spectrum over the full wavelength and flux range. The three insets show zooms around faint spectral features.  

\begin{figure}
\includegraphics[width=\columnwidth]{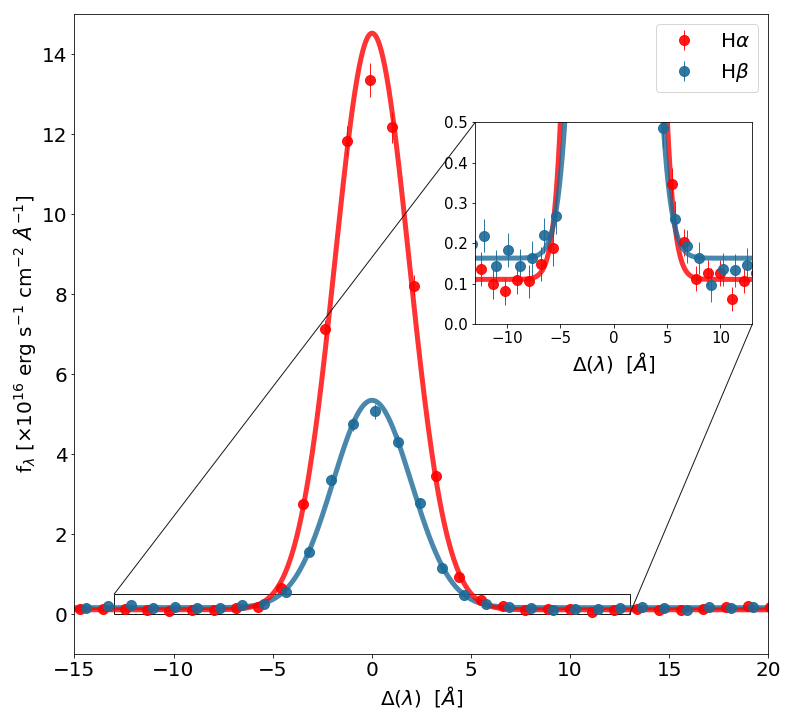}
\caption{Full emission line profiles  for both the  \ha\ and \hb\ lines. The inset zooms in on  the continuum around the lines. The solid curve indicate  the best-fit Gaussian profiles to the emission lines. 
 \label{fig:fullprofiles}} 
\end{figure}

Fluxes of isolated emission line were computed via direct integration of the continuum--subtracted line profiles. The continuum was estimated by computing the median flux over 5\AA\ on each side of the emission line.
This approach was chosen because the  line profiles, when  measured at high signal-to-noise, were found to  deviate from  perfect Gaussian functions in the core region. This discrepancy is shown in  Figure~\ref{fig:fullprofiles}, where we present the full line profiles for both the \ha\ and \hb\ emission lines together with the best fit Gaussian functions\footnote{The fits were performed  using the  {\fontfamily{qcr}\selectfont  optimize.curve\_fit} algorithm within the SciPy libraries \citep{scipy}} (solid curves). However even when computing the line ratios using Gaussian integration, the result remains unchanged. The inset zooms in on the continuum to allow the reader a visual assessment of the data quality. 
Given the large EW of the emission lines, the  error budget on  the total line  flux is dominated by the flux calibration uncertainty.  
The flux measurements, relative to \hb, and the corresponding errors are listed in Table~\ref{tab:one}. 
We measure a Balmer \ha/\hb\  of  $2.620\pm0.078$, significantly smaller than the canonical Case~B recombination theory limit, commonly assumed in the analysis of  \hii\ regions. For typical electron temperatures ($T_e = 10^4$K) and electron densities ($n_e = 10^2$ cm$^{-3}$), a completely dust free \hii\ region would have (\ha/\hb)$_{\rm Case B} = 2.863$ \citep[e.g.,][]{1989agna.book.....O}. 

\subsection{Impact of observational effects}
We explored (and rejected) the possibility that the low Balmer ratio is a result of observational effects producing a wavelength dependent bias in the measured fluxes. 

We first consider  atmospheric differential refraction \citep{Filippenko1982} and chromatic seeing \citep{Fried1966,Boyd1978}. Both these effects can introduce a wavelength dependent biases. However, they cause a larger flux loss at the blue end of the spectral range compared to the red end and thus, corrections for these effects would  make the \ha/\hb\ ratio smaller. The differential atmospheric refraction is corrected for via the atmospheric dispersion corrector installed at the secondary mirror of the MMT \citep{Fabricant2008}. 

We estimate the contribution of chromatic seeing following the recent work of \citet{Herenz2017}. These authors use Integral Field Spectroscopic (IFU) data to characterize the wavelength dependency of the seeing FWHM. Using their best fit linear fit \citep[Table~2 in][]{Herenz2017}, we find that the seeing is $\approx$ 9\% larger at 5200\AA\ compared to 7000\AA\, (the observed wavelengths of \hb\ and \ha\ respectively). Accordingly, given the 1\farcs0 seeing during the  observations, the energy enclosed by the 1\farcs5 Hectospec fibers is $\approx$ 7\% smaller at 5200\AA\ compared to 7000\AA. These calculations are valid for point sources.  
Given the uncertainty on the real galaxy profile, we do not apply this correction to the observed line ratio. We however caution the reader that the real \ha/\hb\  could be as low as 2.43, as a consequence of chromatic seeing.

As mentioned above, our observations were conducted without an order-sorting filter to remove second order blue light.  This phenomenon only affects spectrophotometry of the stellar continuum radiation, unless by chance a blue emission line would fall at exactly the same detector coordinate as \ha, and we have verified that this is not the case.  
 At the observed wavelength of the \ha \ emission line ($\sim 7020$\AA) the second order contamination of stellar continuum  is negligible, due the  combination of the second order blaze throughput curve falling off and the fiber throughput also decreasing at 3500\AA\  (Fabricant, private communication).  To test  this,  we compared the sensitivity functions inferred  from  the three standard stars, observed over  the course of three nights. These stars span a broad range of  colors,  with m$_{3500}-$m$_{7000}$ varying from $-$1.4 to $+$1.4, AB magnitudes. If second order contamination  at 7000\AA\ were important, we would  expect to see a  correlation between relative differences in the sensitivity curves and the stellar colors.  We do not see a measurable trend with stellar color, confirming that  the effect is not important at the accuracy of the calibration.

We also examine the possible contribution of underlying stellar absorption at the wavelength of the Balmer \ha\ and \hb\ lines.   We see no indications of absorption. Given the large EWs of the Balmer emission lines (i.e., EW[\hb]$=165\pm 14 $\AA), and the extremely hard ionizing radiation field implying very hot stars (Figure~\ref{fig:one}), this is not surprising. The underlying absorption correction for a stellar age of a few Myrs is smaller than 3\AA\ at \hb\ \citep[][]{1999ApJS..125..489G} and it is larger at \hb\ than \ha. Thus, similarly to the effects of atmospheric refraction,  more flux would be lost to \hb\  than \ha. Correcting for stellar absorption would then result in an even lower \bal\ ratio.

Finally,  we applied a correction for Milky Way reddening, using the \cite{Schlafly2011} recalibration of \citet{schlegel1998} extinction maps. Given the low resolution of these maps, we consider the possibility that this correction is overestimated. The most conservative assumption would be no MW extinction at the position of SXDF308. An E(B-V)=0, however,  would increase the \bal\ ratio to 2.649$\pm$0.078, still inconsistent with the typical Case~B recombination theory value of 2.86 at $\approx$ 2.7$\sigma$.

To conclude, observational effects are not able to explain the low value of the \bal\ ratio.

\section{Physical explanation of the low \bal\ ratio}
\label{sec:interpretation}
In order to assess the implications of the measured Balmer ratio, we need to first derive the physical conditions of the nebula. This is done in Section~\ref{sec:Te_ne}. We then review the values expected in textbook Case~B and Case~A limits (Section~\ref{sec:caseAB}), and move on to explore other mechanisms that can impact the emissivity of the Balmer lines. The Case~C limit is discussed in section~\ref{sec:caseC}, and Balmer self absorption in Section~\ref{sec:balmer}. As we argue below, we conclude that Balmer self absorption in non-spherically symmetric gas is the favored interpretation. 

\subsection{Estimate of $T_e$, $n_e$}
\label{sec:Te_ne}

The electron temperature ($T_e$) in the $O^{2+}$ region can be estimated directly using the ratio between the intensity of the $^1S_0 \rightarrow\, ^1D_2$ 4363\AA\ emission line and the combined intensities of the $^1D_2 \rightarrow\, ^2P_2$ 4959\AA\ and $^1D_2 \rightarrow\, ^2P_1$ 5007\AA\ lines. The energies required to collisionally excite the $^1S_0$ and $^1D_2$ states in the O$^{2+}$ ion are 5.4 and 2.5eV respectively. Thus, in the limit of low electron density ($n_e < 10^3$cm$^{-3}$), as  $T_e$ increases, the relative population of ions in the  $^1S_0$ state increases, resulting in a relative brightening of the 4363\AA\ line. 
We use  the {\fontfamily{qcr}\selectfont  PyNeb}  software \citep{PyNeb} to compute the electron temperature, using the atomic data of \citet{FroeseFischer2004}. We compute an electron temperature of  $T_e(O^{2+})=15380 \pm 850$K. This electron temperature is higher than typical values observed in galactic \hii\ regions, but is consistent with the temperatures measured in the most metal poor galaxies \citep[e.g.,][]{Skillman2013}. We cannot compute the temperature of the low ionization zone directly, as the S/N of the spectrum is not sufficient to measure lines past rest--frame 7100\AA, and the \nii$\lambda$5755 is not detected, likely because of the low nitrogen abundance. Thus, we follow \citet{Skillman2013} and use the \citet{Pagel1992} relation between $T_e(O^+)$ and $T_e(O^{2+})$\footnote{Defining $t_3 = \frac{10^4}{T_e(O^{2+})}$ and  $t_2 = \frac{10^4}{T_e(O^{+})}$, then $t_2^{-1} = 0.5(t_3^{-1} +0.8)$. This relation depends on the photoionization modeling, and \citet{Pagel1992} used the calculations by \citet{Stasinska1982} }. 
Similarly, we used the relationship derived by \citet[][see footnote \footnote{$t(S^{2+})=0.83t_3 +0.17$}]{Garnett1992} to derive the temperature of the S$^{2+}$ region. The computed temperatures are tabulated in Table~\ref{tab:two}.

Finally, we estimated an electron density of $n_e(S^+) = 10^{+200}_{-10}$cm$^{-3}$ using the \sii$\lambda\lambda$6717,6731 doublet. In addition to the \sii\ doublet, we also detect the \ariv$\lambda\lambda$4712,4740 doublet, originating in high excitation gas. 
The ratio between the two emission lines ($R1\equiv I_{4712}/I_{4740}$) can be used to derive the electron density, following \citet{keenan1997}. At the resolution of the MMT spectrum, the \ariv$\lambda$4712 line is contaminated by the \hei~4713 line. We estimated the \hei~4713 line flux from the \hei~4471 line using the emission line ratios presented in  \citet{Porter2012}. 
The corrected ratio $R1$=0.7$\pm$0.4 only allows us to place an upper limit on the $n_e(Ar^{3+})<10^3$cm$^{-3}$.

\subsection{Case~B and Case~A limit}
\label{sec:caseAB}
\begin{figure}
\includegraphics[width=\columnwidth]{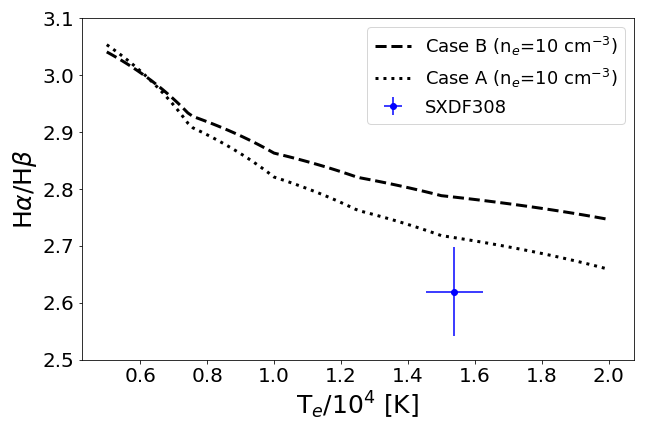}
\caption{\ha/\hb\ ratio as a function of electron temperature, computed for two case limits (A, B), assuming a gas electron density of 10 cm$^{-3}$.\label{fig:two}}
\end{figure}

Emission line spectra from star-forming galaxies are interpreted within the successful Case~B recombination limit theory, which  describes the limit in which all recombination photons in the Lyman series are scattered in a spherically symmetric medium often enough to be eventually degraded into Balmer and higher series photons and either two-photon continuum or \lya. 
We used  {\fontfamily{qcr}\selectfont  PyNeb}  to compute the expected Case~B Balmer ratio, given the gas temperature and density measured in Section~\ref{sec:Te_ne}. We find  (\bal)$_{\rm Case B} = 2.79$, 2.2$\sigma$ higher than the observed value. 
The opposite limit from Case~B is Case~A, where the \hi\ column density is low enough for the gas to be optically thin in all recombination lines, including those of the Lyman series. In this limit, the emissivities of Balmer photons are reduced, and the predicted \ha/\hb\ ratio is lower than the corresponding Case~B value. Figure~\ref{fig:two} shows the variation of the Balmer ratio with electron temperature, computed for  $n_e=10$ cm$^{-3}$ in the Case~A and Case~B limits.  The curve does not change appreciably for $n_e=100$~cm$^{-3}$, at the measured electron temperature. Although the (\ha/\hb)$_{\rm Case A}$ decreases more rapidly with temperature than (\ha/\hb)$_{\rm Case B}$, it never reaches the value observed in SXDF308, although it is  consistent (within 1.5$\sigma$) with it.
 
\subsection{Case-C limit}
\label{sec:caseC}
It is instructive to consider other physical mechanisms that can affect the intrinsic value of the  \ha/\hb\ ratios and change it  from the Case~B limit. One possible mechanism, important for optically thin gas, is the contribution to the Balmer line emission from reprocessed continuum photons. This mechanism, known as Case~C, was first discussed by \citet{Baker1938} and more recently by \citet{Ferland1999} and \citet{Luridiana2009}. Case~C relaxes the assumption, implicit in the Case~A and Case~B calculations, that excited levels can only be populated by radiative recombinations and cascades from higher levels, and includes the possibility that they can be effectively excited via fluorescence, i.e., through the absorption of {\it  continuum} photons at the wavelength of the Lyman lines with subsequent decay into n$\geq 3$ levels. Under Case~C, the resulting emission line spectrum from the nebula depends not only on its physical conditions (i.e., electron temperature and density), but also on the detailed shape of the continuum at energies between 1 Ryd and \lya.  \citet{Ferland1999} and \citet{Luridiana2009} show that continuum pumping contributes substantially to the \ha/\hb\ ratio (compared to Case~A) in the most internal regions of  ionized clouds. 
 
\begin{figure}
\includegraphics[width=\columnwidth]{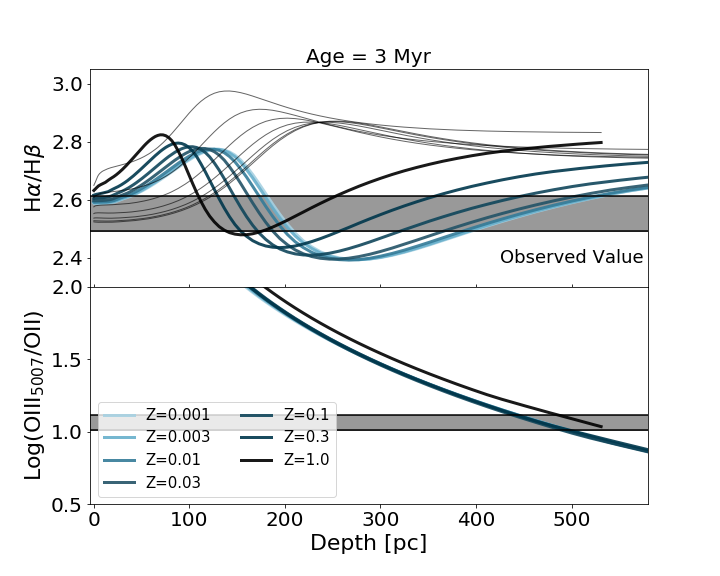}
\caption{{\it Top panel:} Predicted \ha/\hb\ ratio as a function of depth, computed for a 3Myr burst with different  metallicities (colored lines, see legend for details). Also shown are models in which the continuum radiative pumping of the hydrogen Lyman lines was turned off (thin grey lines). {\it Bottom panel:} same as top panel, but for the \oiii/\oii\ ratio (curves with metallicity smaller than $Z_{\odot}$ lie very close to each other).  \label{fig:cloudy}} 
\end{figure}

In order to explore the viability of this interpretation we computed photoionization models using Cloudy version 17.01 \citep{cloudy17}. The photoionization models presented in Figure~\ref{fig:cloudy} assume a dustless spherical cloud with a inner radius of 1pc and a constant hydrogen density of $n_{\rm H}$=10 cm$^{-3}$.  
 We modeled the ionizing source using  a single burst stellar population models created with  Starburst~99 \citep[Version 7.0.0][]{Leitherer2014}. We run the calculations of the stellar population synthesis models using the Geneva stellar tracks (both with and  without rotation), assuming  a  single  stellar  metallicity of $Z= 0.008$, and a \citet{Kroupa2001} initial mass function.  

In addition to the \ha/\hb\ ratio, we also computed the expected O$_{32}$ ratio and \hb\ luminosity as a function of the cloud's depth. We know that SXDF308 shows an unusually high O$_{32}$ ratio (O$_{32} = 11.80 \pm 0.74$), so any photoionization model needs to be able to reproduce the high excitation state of oxygen as well. 
The varying line ratios as a function of cloud's depth (computed using Cloudy's cumulative intensities) are presented in Figure~\ref{fig:cloudy} for a burst age of three million years. Older bursts were not able to reproduce the observed ranges of line ratios, indicated in the Figure with horizontal grey areas. Lines of different color show models with varying total gas metallicity between Z$_{\rm gas}$=1 and 0.001Z$_{\odot}$, as indicated in the Figure. All models show similar trends in the \ha/\hb\ ratio with depth, which is due to the varying contribution of fluorescent emission in the two emission lines.  \ha\ and \hb\ fluorescent photons are mainly produced after the absorption of Ly$\beta$ and Ly$\gamma$ continuum photons, which are abundant in the illuminated part of the nebula. Because of the larger  Ly$\beta$ optical depth compared to Ly$\gamma$, the \ha\ emissivity peaks at smaller depth than \hb, resulting in the observed variation with depth.  Moving deeper into the cloud, Lyman line photons are consumed and the \ha/\hb\ ratio approaches the Case~B limit. The thin grey lines in the top panel of Figure~\ref{fig:cloudy} show, for reference, a set of identical calculations where the continuum radiative pumping of the hydrogen Lyman lines was turned off. All line ratios converge at large depth. Decreasing the gas metallicity pushes the curves to the right, i.e., the same value of  \ha/\hb\ is observed deeper in the cloud. The gas metallicity mainly affects the cloud electron temperature, going from T$_{e}$ of 12kK to 15kK for Z$_{\rm gas}$ of 0.1 and 0.025Z$_{\odot}$, respectively, resulting in a change of the recombination-only line ratio. Indeed, the curves are shifted to the right even when continuum pumping is turned off.

All explored models are able to reproduce the observed value of the hydrogen emission line ratio in the depth range between approximately 100 and 500 pc, depending on the gas metallicity. The situation is, however, different when the ratio of the oxygen lines is considered. The O$_{32}$ ratio as a function of cloud depth is shown in the bottom panel of Figure~\ref{fig:cloudy}. The lines represent the same models as in the top panel, although we do not show the models with continuum pumping turned off, since it does not affect the O$_{32}$ values. As expected, the oxygen ionization state and the resulting O$_{32}$ line ratio is very sensitive to the position in the cloud, and depends very little on the gas metallicity. The combined constraints from the Balmer and oxygen line ratios limit the viable  range of gas metallicities to $Z< 0.1Z_{\odot}$.

In addition to line ratios, the model needs to reproduce the absolute luminosity of the Balmer lines. In order to produce line ratios within the observed ranges, the 3~Myr starburst was normalized to a total number of ionizing photons (Q) of $\log{Q}=52$. The Cloudy total  \hb\ line luminosity for this optically thin cloud is, however, a factor of $\approx 10 $ smaller than the observed 3$\times 10^{40}$ erg s$^{-1}$.  Although it is possible to obtain a  \hb\ luminosity consistent with the  observed value, these models would require a near UV flux many orders of magnitudes larger than the observed one.  Invoking absorption of dust is not a viable option given the observed line ratios. Thus, although promising, the fluorescent mechanism cannot be uniquely responsible for  the observed line ratios and luminosities. 

\subsection{Balmer self--absorption}
\label{sec:balmer}
Another mechanism that can affect the relative intensity of the Balmer lines is self-absorption of  Balmer photons. This effect is important if the gas' optical depth to Balmer photons is not negligible and requires a large population of hydrogen atoms in the $n=2$ energy level. 

Figure~\ref{fig:balmerratios} shows the position of SXDF308 in the \ha/\hb\ versus \hg/\hb\ ratio diagram, together with the values of the emission line ratios predicted for Balmer self-absorption \citep[from][]{Cox1969}.  Specifically, each track shows the calculations for constant \lya\ optical depth (labeled in each curve) and varying \ha\ optical depth  (on each track we mark the positions for $\tau_{H\alpha}= 5$ and 10).  The values of the ratios expected for dust-free Case~B recombination theory are also indicated (thin black  lines). When Balmer self-absorption is important, \hab\ smaller than the Case~B limit is predicted for a medium with a low optical depth to \lya\ photons  ($\log{\tau_{Ly\alpha}}\sim 2.6 $). This is because  in the limit of low optical depth to the \lya\ line, \ha\ photons scattered in the neutral gas are effectively converted into \lyb\ photons and escape. 
In this (unrealistic) situation, however, the medium would be very thin to ionizing radiation, with a $\tau_{LyC} \sim 0.04$, making the nebula substantially fainter in recombination lines.

In the more realistic scenario of  a gas is optically thick to  Lyman line photons, as the  \ha\ optical depth increases  the \ha/\hb\  ratio also  increases, because \ha\ photons are simply scattered in the medium, while \hb\ photons get converted into \ha$+$Pa$\alpha$. In this regime, however, the geometry of the scattering medium becomes important, and the effects of radiative transport of Balmer photons cannot be neglected. If the scattering gas is asymmetric and preferential directions exist that offer lower total optical depth to Balmer photons, the resulting \ha/\hb\ ratio depends on the specific line of sight into the nebula and low values of the \ha/\hb\ ratio can be allowed. This would be the case if, for example, the gas were organized in a flattened geometry, oriented edge--on with respect to the line of sight. In this case, \ha\ photons would preferentially escape along the direction perpendicular to the disk (i.e., perpendicular to the line of sight), thus effectively reducing the \ha\ output toward the observer and decreasing the  \ha/\hb\ ratio. Detailed radiative transfer calculations are being performed to quantify this effect and will be presented in a forthcoming paper.

\section{Discussion}
\label{sec:discussion}
\begin{figure}
\includegraphics[width=\columnwidth]{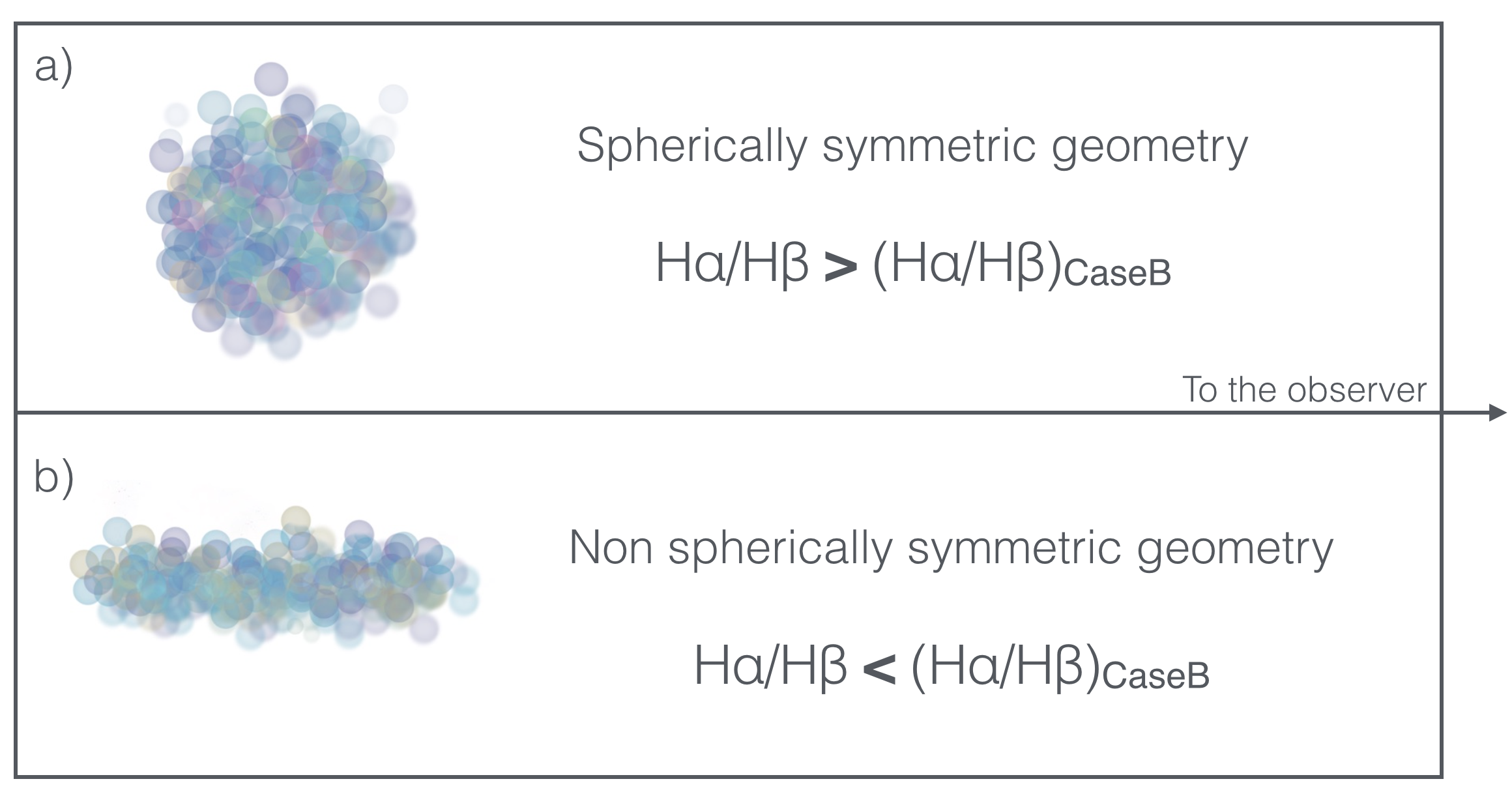}
\caption{Schematic diagram of the scenario proposed in this paper. \ha/\hb\ Balmer line ratios smaller than predicted for Case~B recombination theory are the result of preferential scattering of \ha\ photons out of the line of sight, in  galaxies where the scattering gas shows an  asymmetric geometry (panel b). In galaxies with  a symmetric gas distribution both the \ha/\hb\ and \hg/\hb\ ratios would be larger than Case~B, indicating the effect of Balmer self-absorption.  \label{fig:schematic}} 
\end{figure}

Section~\ref{sec:interpretation} presents evidence that the canonical Case~B limit interpretation of \hii\ region emission line spectra fails for SXDF308. We suggest  that Balmer self-absorption and scattering of Balmer line photons may be important mechanisms to regulate the intensity of the Balmer emission lines. 
For SXDF308, the low Balmer \hab\ ratio can be explained if the geometry of the medium responsible for scattering the Balmer photons is not spherically symmetric, but rather is similar to a disk oriented edge-on with respect to the line of sight. In this configuration, \ha\ photons would preferentially be scattered off the line of sight, explaining the observed values.  Figure~\ref{fig:schematic} shows a schematic diagram of the proposed scenario. Galaxies with low \ha/\hb\ Balmer line ratios are the result of preferential scattering of \ha\ photons out of the line of sight, in galaxies where the scattering gas is characterized by an  asymmetric geometry (panel b in Figure~\ref{fig:schematic}). Galaxies with symmetric gas distribution would show both  \ha/\hb\ and\ \hg/\hb\ ratios  larger than Case~B, as a result of Balmer self-absorption (panel a).   

If our interpretation is correct we would expect the effects
of Balmer self-absorption to be at play in other galaxies. However, given only the \ha, \hb, \hg\ observations as we have here, these galaxies would only be identifiable under certain conditions: that they have a polar asymmetry in the column densities, are oriented edge-on, and have very small/negligible nebular reddening. Conditions other than this would all result in line ratios consistent with reddened gas, and they would mostly go unnoticed.
Therefore we need to explore whether or not evidence exists  in other galaxies that Balmer self-absorption is not negligible.  In general, we would expect that radiative transport effects would  not always work in favor of \hb\ as in the case of SXDF308. Consequently, we should see a population of galaxies with {\it both} the \hab\  {\it and}  the \hg/\hb\ ratios larger than Case~B. Clearly, a large value of the  \hab\ alone cannot  be uniquely interpreted as the result of Balmer self-absorption, as dust extinction would also go in the same direction. However, the two mechanisms have a different effect on the \hg/\hb\ ratio: while dust extinction {\it decreases}  the \hg/\hb, Balmer self-absorption {\it increases} it \citep{1989agna.book.....O}. 

\begin{figure}
\includegraphics[width=\columnwidth]{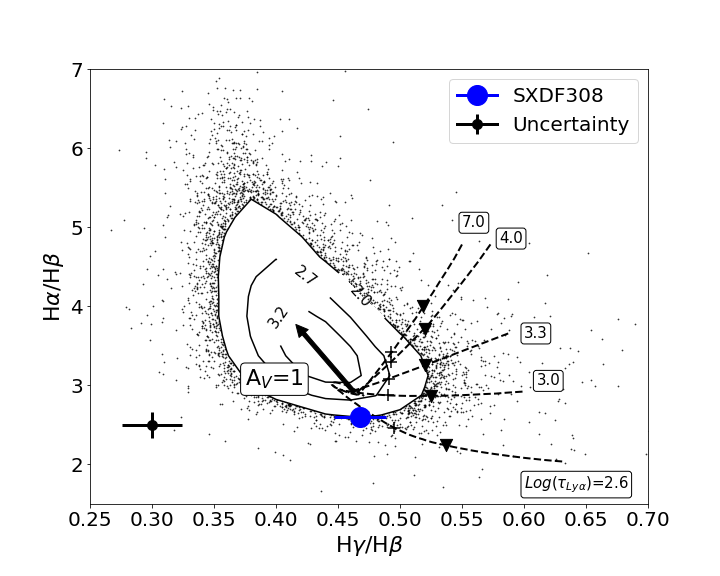}
\caption{ \ha/\hb\ as a function of the \hg/\hb\ ratio for star-forming galaxies in the SDSS survey. The contours indicated the number of galaxies per bin of \ha/\hb\ and \hg/\hb\  ratios, starting from a density of 100 galaxies. We plot individual points for lower densities of galaxies. The dashed lines show the predicted ratios for varying \lya\ optical depth (labeling each track), and \ha\ optical depth (increasing along each track). Along each track, we mark the position of $\tau_{H\alpha}=5$ and 10, with  plus signs and triangles respectively. Galaxies in the upper left quadrant of the plot can easily  be explained by dust extinction, while galaxies in other parts of the diagram cannot.  The figure shows that there is a population of galaxies with line ratios inconsistent with extinguished Case~B values,  but that can be reproduced with Balmer self-absorption and a range of optical depth to \ha\ and \lya\ photons.  \label{fig:balmerratios}} 
\end{figure}

 To  test if such  a population of galaxies exists we explore the ratios of Balmer lines  for star-forming galaxies in the SDSS survey (see Figure~\ref{fig:balmerratios}). We use the publicly available\footnote{Available from https://www.sdss.org/dr14/spectro/galaxy\_mpajhu/} MPA-JHU Data Release~8 catalogue, developed by the Max Planck Institute for Astrophysics and the Johns Hopkins University groups \citep{Kauffmann2003,Brinchmann2004}.  The measurement of the nebular emission lines  accounts for the galaxy stellar continuum and stellar absorption at the Balmer lines as  explained in,  e.g.,\citet[][]{Tremonti2004}. To realistically account for uncertainties in the stellar-continuum  subtraction, we increased the emission line errors by factors determined by comparisons of duplicate observations within the SDSS sample. These factors, listed on the catalog  website\footnote{http://www.mpa-garching. mpg.de/SDSS/}, amount to 1.882  and 2.473 for the \hb\ and \ha\ lines, respectively.  Additionally, \citet{Groves2012MNRAS} report that the stellar absorption EW of \hb\ is  systematically underestimated by 0.35\AA\ in the MPA-JHU catalog. We apply the recommended correction to the \hb\ emission line fluxes\footnote{$F_{\rm H\beta}^{\rm corrected} =F_{\rm H \beta}^{\rm SDSS} + 0.35/{\rm EW(H_{\beta})}  F_{\rm H \beta}^{\rm SDSS} $}.

\begin{figure*}
\includegraphics[width=\textwidth]{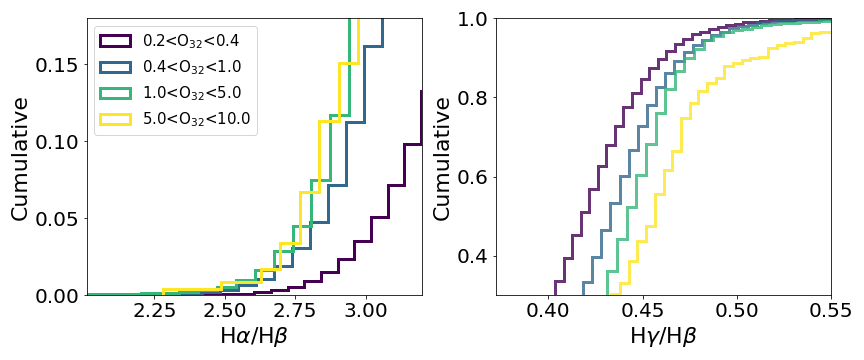}
\caption{Cumulative distributions of \hab\ (left) and H$\gamma$/H$\beta$ (right) line ratios  binned according to the O$_{32}$ line ratio.  The fraction  of galaxies with \hab(H$\gamma$/H$\beta$) lower (higher) than Case~B recombination theory increases with the O$_{32}$ ratio\label{fig:cum}.} 
\end{figure*}

In  Figure~\ref{fig:balmerratios} we only include SDSS galaxies with emission line fluxes measured with a signal-to-noise ratio greater than 15 in all lines,  and exclude Active Galactic Nuclei (AGN) using the spectral classification based on the \citet[][BPT]{1981PASP...93....5B} diagram, with the separation line between AGN and star-forming galaxies of \citet{Kauffmann2003agn}. The black point in the lower left of the panel shows the median uncertainty as shown by the error bars. To avoid saturation of the density of point, we show isodensity contours where the density of SDSS galaxies is larger than  100 galaxies per bin of \ha/\hb\ and \hg/\hb\ (corresponding to the outermost contour), and individual points where the density is lower. 

The arrow in Figure~\ref{fig:balmerratios} shows the direction in which a point would move for an extinction of one magnitude in the $V$ band, for a \citet{Calzetti2000} extinction law and the assumption of a uniform screen geometry in front of a point source. Changing the dust distribution to a clumpy screen $+$ extended source would still not help explaining galaxies with \ha/\hb\ and \hg/\hb\ smaller and larger, respectively,  than those predicted for Case~B recombination theory  \citep[e.g.,][]{scarlata2009}.  Clearly, there exist a population of galaxies on the right of the Case~B line at \hg/\hb$=0.466$ with line ratios that cannot be explained with dust extinction. Instead, these line ratios are  consistent with those predicted by calculations including Balmer self-absorption. The scenario that emerges from these observations is that  the effects of self-absorption of the Balmer lines in galaxies may not be negligible as commonly assumed.  

The consequences of this conclusion are important:

\noindent First,  analysis of emission line spectra  that proceeds from the assumption that Case~B recombination limit holds will be erroneous. Dust extinction correction to galaxy spectra are usually computed using only the Balmer \ha/\hb\ ratio, under the assumption that a value greater than Case~B is uniquely due to dust. This assumption will cause an over-correction of the observed spectra, and thus an overestimate of quantities such as the intrinsic star-formation rate, galaxy ionizing power, {\it etc}. Adding the intensities of Balmer lines originating from $n>3$, as demonstrated in Figure~\ref{fig:balmerratios}, can help in identifying those objects where absorption of Balmer line photons is important. 

\noindent
Second, if, in addition to Balmer self-absorption,  dust  is also present in a galaxy, then the line ratios will move to the region of the diagram allowed by Case~B $+$ dust, and these galaxies could only be identified  with the addition of other lines, such as, e.g., those of the Paschen series. {Emission lines of helium could also be used to identify departure from Case~B due to radiative transfer effects in hydrogen. At the signal-to-noise of our  spectrum, the \hei\ lines are consistent with no absorption by dust. It is also worth mentioning that radiative transfer of Balmer photons will also alter other ratios involving Balmer  lines, like the R23 Pagel's index \citep{Pagel1992}, and  the  indices used in the BPT diagrams.   For large \ha\ optical depths, galaxies will move upward on the BPT, toward high \oiii/\hb\  ratios.  }

\noindent Third, the possibility that in some galaxies hydrogen atoms are  in the excited state implies that it is possible to ionize them with photons more energetic than the Balmer ionization energy limit of 3.4eV (including \lya\ photons), rather than Lyman continuum photons. This possibility, in turn, implies that the luminosity of the Balmer lines is not a direct tracer of Lyman continuum photons, as assumed in the Case~B limit. More in general, all calculations relying on the intrinsic intensity of the Balmer lines will be affected, including, e.g.,  element abundances.   

If Balmer self-absorption is   the  correct explanation for the anomalous Balmer line ratios, there has to be a population of hydrogen atoms in the excited $n=2$ state. 
\citet{Purcell1952} and \citet{Pottasch1960} suggest that trapping of \lya\ photons can be responsible of producing a sizable population of excited \hi. Although \lya\ photons can be easily destroyed by dust \citep[e.g.,][]{Ershov1987,Dennison2005,Hayes2015,Dijkstra2016}, SXDF308 in particular, and extreme emission line galaxies  in general \citep{Henry2015}, have a modest dust content, implying a low destruction rate of \lya\  photons.  Detailed calculations are beyond the scope of this paper, however, an order of  magnitude analysis shows that the proposed  mechanism is indeed possible. In order to decrease the \hab\ ratio by scattering $\sim$10\% of \ha\ photons off  the line of sight, we need $\tau_{H\alpha}\sim 0.1$. According to \cite{Ferland1979} $\tau_{H\alpha}\sim  4\times 10^{-7} \langle N_{Ly\alpha} \rangle  Q(H)_{52} R^{-2}$, where $\langle N_{Ly\alpha} \rangle $ is the average number of  \lya\ scattering, $Q(H)_{52}$ is the number of hydrogen ionizing photons per second in units of $10^{52}$ s$^{-1}$ and $R$ is the distance of the scattering medium to the source in pc. The observed \ha\ luminosity of $\approx 10^{41}$ erg s$^{-1}$ translates into a $Q(H)_{52} \gtrsim 10$. Assuming a size of $\sim 10$pc, $\tau_{H\alpha}\sim  0.1$ implies $\langle N_{Ly\alpha} \rangle \sim 10^6$, ensuring that \lya\ pumping of the excited state is effective. 

Extreme emission line galaxies such as SXDF308 and the Green Peas are the only galaxy population to consistently 
show \lya\ in emission \citep[i.e., more than 90\%,][]{Henry2015,Yang2017b}. They are therefore an obvious population in 
which \lya\ pumping may be at play. To further examine this idea we explore the dependency of the Balmer ratios on the 
$O_{32}$ line ratio:  the most extreme values of O$_{32}$, similar to 
SXDF308, are a common characteristic of extreme emission line galaxies. In Figure~\ref{fig:cum} we show the distributions of \ha/\hb\ (left) and H$\gamma$/H$\beta$ (right) line ratios binned  according to O$_{32}$. These panels show a clear trend with a larger fraction of galaxies having lower(higher) 
\hab(H$\gamma$/H$\beta$) value than the Case~B. 

It is important to exclude the possibility that the trends observed in Figure~\ref{fig:cum} are  a consequence of systematic biases introduced by the SDSS data reduction pipeline. Such effects can be important as demonstrated in, e.g., \citet{Lan2018} and \citet{Rodriguez2020}. Specifically, \citet{Lan2018} shows that the flux calibration of the SDSS spectra can result in three main spurious spectral features: $i)$ wiggles in the $4500<\lambda<7000$\AA\ wavelength range (due to mismatches between stellar templates and observations)  that can introduce additional random uncertainty to the flux calibration at the level of at most 1\%; $ii)$ a strong absorption at the wavelength of  Calcium H\&K ($\lambda\lambda$3934,3969) and Sodium~D ($\lambda\lambda$5891,5897), due to absorption by the Milky Way interstellar medium;  and finally $iii)$ a strong absorption line at the wavelength of \ha\, due to a problem in the masking of the calibration spectra around this wavelength. The first effect is not an issue because we  consider galaxies at a range of redshifts. This effect therefore introduces random rather than systematic uncertainties in the flux calibration (i.e., for some galaxies a line will fall on a minimum of the wiggle, and for others it will fall on a maximum). The latter two points, however, could be an issue if the sample of SDSS galaxies we considered has any of the lines of interests (\ha, \hb, \hg, \oiii, and \oii) falling close to some of the absorption features. We therefore computed Figure~\ref{fig:cum} excluding all objects at those redshifts where any of the lines could be compromised.

The trends presented in Figure~\ref{fig:cum} strongly suggests that there is a link between the galaxies ionization state and the observed Balmer ratios, possibly indicating that a significant population of galaxies with strong Balmer self absorption could exist and hide among the ostensibly dust--reddened population.

\section{Conclusions}
\label{sec:conclusions}
We  presented the emission line spectrum of SXDF308, an extreme emission line galaxy showing a \ha/\hb\ Balmer line ratio significantly lower that the value predicted for dust free Case~B recombination theory. We ruled out the possibility that this ratio is due to systematic errors introduced by observational effects (e.g., wavelength dependent seeing, atmospheric dispersion). We discussed various mechanisms that can alter the intrinsic ratios of the Balmer emission lines, and exclude both Case~A and Case~C as explanation of the observed spectrum. The mechanisms that best reproduces the observations of SXDF308 invokes resonant scattering of \ha\ photons, in a non spherically-symmetric geometry that would preferentially scatter \ha\ photons  away from the line of sight. Basic calculations show that some extreme starburst galaxies produce sufficient \lya\  to effectively populate the n=2 level of hydrogen. 

Using data from the Sloan Digital Sky Survey, we show that Balmer self-absorption may be relevant more generally than just in this galaxy, with important implications for the interpretation of galaxy spectra.  If the Case~B limit is not valid,   galaxies' physical properties computed from the intensity of the Balmer lines will be  incorrect, if Case~B is assumed. These properties include, but are not limited to, the dust extinction, the intrinsic ionizing power, and the intrinsic star-formation rate. 

SXDF308, together with the other examples among extreme emission line galaxies  stand out and are identified as problematic because of their "unphysical" low Balmer \ha/\hb\ ratio,  which in our explanation, is likely due to the particular geometry of the scattering gas. In general, though, galaxies for which radiative transfer effects of Balmer photons are important cannot be identified on the basis of this ratio alone, as, in the more general predictions, their values overlaps with the range expected for Case~B plus extinction by dust. The addition of higher order Balmer lines and Paschen lines can help identifying these cases. More detailed modeling and additional observations are required to understand the impact of Balmer self-absorption to the interpretation of galaxies' spectra. 

\section*{Acknowledgements}
CS and VM acknowledge the support from Jet Propulsion Laboratory under the grant award \#RSA-1516084. VM also acknowledges support from the University of Minnesota Doctoral Dissertation Fellowship 2016-17.
M.H. acknowledges the support of the Swedish Research Council, Vetenskapsr{\aa}det and the Swedish National Space Board (SNSB), and is fellow of the Knut and Alice Wallenberg Foundation.
This work made use of the following software packages: ipython \citep{ipython}, numpy \citep{numpy}, matplotlib \citep{matplotlib}, astropy \citep{astropy}.



\bibliographystyle{mnras} 
\bibliography{myreferences}{}



\appendix

\section{Assessing systematic uncertainties}
A number of factors can limit the accuracy to  which line ratios can be measured.  In this Appendix we assess the impact  of the most important contributors to the flux uncertainty.  

\subsection{Atmospheric absorption}
Atmospheric absorption bands become increasingly important for  wavelengths above 6500\AA, and contribute to add substantial uncertainty to the flux of narrow emission lines \cite[][]{Stevenson1994,Fabricant2008}. These bands are formed by dozens of intrinsically narrow absorption lines that get blurred together in low resolution  spectroscopy, masking the exact nature of the absorption. \citet[][]{Stevenson1994}  caution that when the intrinsic profile of an object's spectroscopic feature has a  width  comparable or smaller than the width of the telluric features ($\sigma < 40$ km  s$^{-1}$),  the standard correction procedure to divide by an observed telluric reference spectrum may fail, depending on the exact relative position of the feature of interest and  the narrow  atmospheric absorption lines.      

\begin{figure}
\includegraphics[width=\columnwidth]{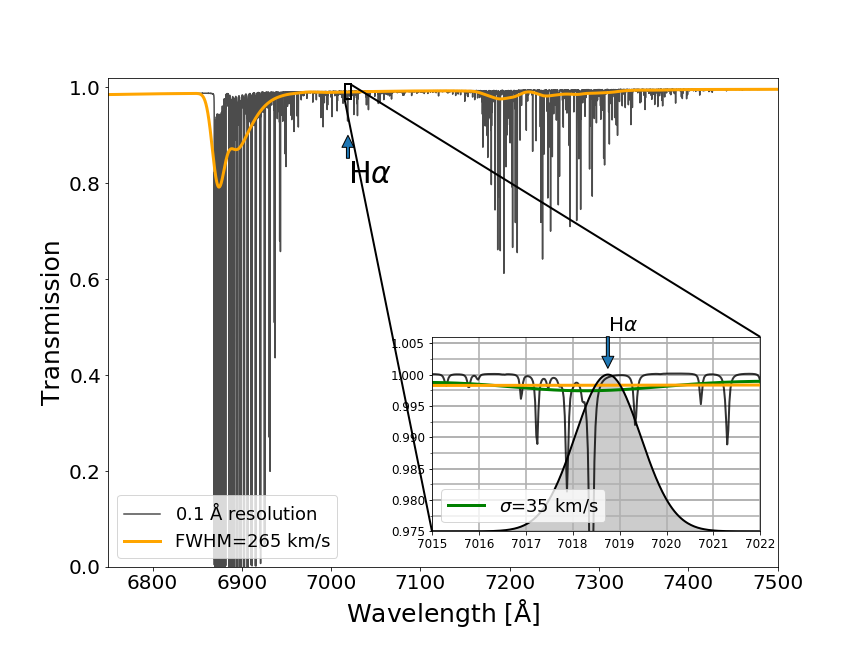}
\caption{The \ha\ line falls in a spectral region clean of strong telluric absorption features. The atmospheric transmission is shown at the intrinsic resolution of the \citep{Noll2012} library (0.1 \AA) and at the resolution of the observations (6.2 \AA, or 265 km  s$^{-1}$). The inset zooms in around the  position of the \ha\ emission line, to show the difference between  the correction calculated for the observed spectral resolution  and  assuming an intrinsic Gaussian emission  line profile with $\sigma =35$ km s$^{-1}$. This difference is insignificant at the position of  \ha. Given the integrated nature of  the observations, we do not expect the intrinsic profile to be  narrower than $\sigma=35 $km s$^{-1}$  \citep[e.g.,][]{Amorin2012}. \label{fig:transmission}} 
\end{figure}

We checked for this problem by comparing the wavelengths of the SXDF308 \ha\ line, after all appropriate velocity corrections, with the wavelengths of nearby strong telluric absorption lines. The results of this  test are presented in Figure~\ref{fig:transmission}, from where we can see that at the redshift of  SXDF308, the  \ha\ line falls in a  region clear of strong telluric absorption.  In the inset  we  compare  the atmospheric  telluric correction computed for the observed spectral resolution, and  what  we  would obtain assuming a galaxy's Gaussian intrinsic \ha\ profile with a velocity  dispersion  of $\sigma  =  35 $  km s$^{-1}$.  We find that at the wavelength of  the  \ha\ emission line, the difference  is  negligible (below 0.2\%). Given the integrated nature of  the observations, we do not expect the intrinsic line profile to be  narrower than $\sigma =35$ km s$^{-1}$  \citep[e.g.,][]{Amorin2012}.
  
\subsection{Flux calibration}
\label{sec:app2}
The accuracy with which line ratios can be measured in high signal-to-noise spectra is limited by the uncertainty introduced by the flux calibration step \citep[e.g.,][]{Skillman2013}. When only one  exposure of one standard star is available, as in most cases, the flux calibration step introduces a conservative $\approx 2$\% systematic uncertainty to the flux measurement, due to the inherent  uncertainty in the spectra of the standard stars  \citep[][]{Oke1990,Berg2015}. 

\begin{figure}
\includegraphics[width=\columnwidth]{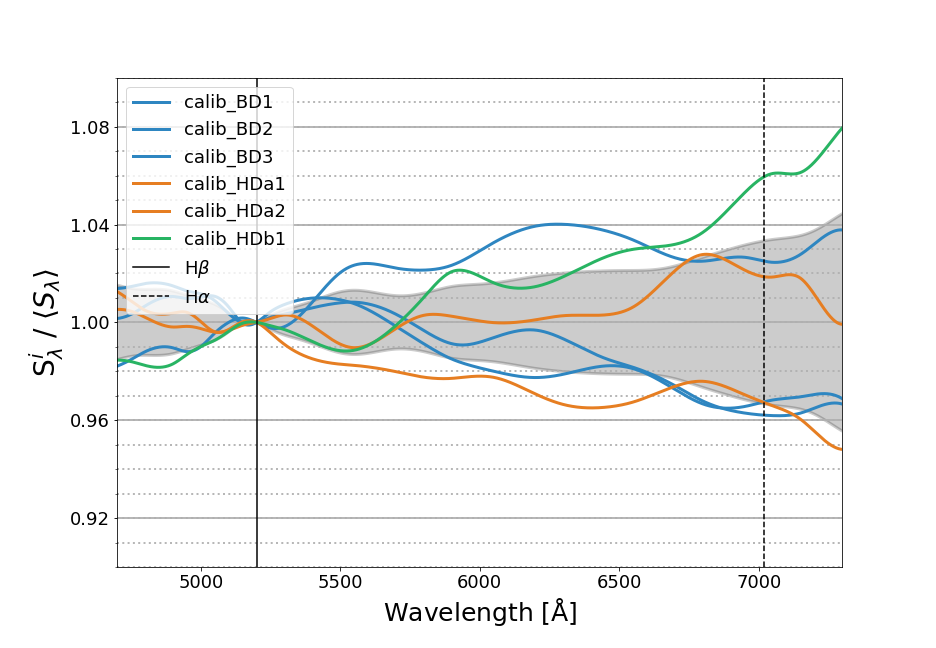}
\caption{Fractional variation among the sensitivity functions obtained from individual standard stars observed multiple times over the course of  the observing run.  Curves of the same color show   sensitivity functions for the same star.  The curves are normalized to the wavelength of the \hb\ emission line, so that the standard  deviation at the wavelength of the observed \ha\ emission line is a direct measure of the uncertainty in the \ha/\hb\ line ratio due to the flux calibration step.  \label{fig:sensitivity}} 
\end{figure}

Our observing strategy included observations of three standard stars observed  multiple times  over the course of three nights. With these observations  we   were able to constrain the relative uncertainty in the \ha/\hb\ line ratio directly.  As explained in Section~\ref{sec:data} the sensitivity function is derived as the average of the sensitivity functions  obtained from all the standard stars. In Figure~\ref{fig:sensitivity} we show the fractional variation between the individual sensitivity functions, normalized at the wavelength of the observed  \hb\  emission line. Curves of the same colors refer to observations of the same stars. The standard  deviation at the wavelength of the observed \ha\ line  is a direct measurement of the uncertainty in the \ha/\hb\ line ratio due to the flux calibration step.

\clearpage

\begin{table}
\caption{MW extinction-corrected emission-line fluxes \label{tab:one}}
\begin{tabular}{cc}\hline
Line		&	I($\lambda$)/I(\hb) \\ \hline
\sii	 $\lambda$6731	&	0.030		$\pm$	0.014 \\
\sii	 $\lambda$6716	&	0.041	$\pm$		0.015 \\
\ha	 $\lambda$6563	&	2.620	$\pm$		0.078 \\
\oiii	 $\lambda$5007	&	6.763	$\pm$		0.141 \\
\oiii $\lambda$4959	&	2.261	$\pm$		0.051 \\
\ariv	 $\lambda$4740	&	0.024	$\pm$		0.008 \\
\ariv+\hei	 $\lambda$4712	&	0.022	$\pm$		0.009 \\
\oiii	 $\lambda$4363	&	0.136	$\pm$		0.014 \\
\hg	 $\lambda$4340	&	0.468	$\pm$		0.021 \\
\hd	 $\lambda$4102	&	0.245	$\pm$		0.015 \\
\oii $\lambda$3727	&	0.573	$\pm$		0.034 \\ 
\hline
\end{tabular}

\end{table}

\begin{table}
\caption{Physical properties for the nebular gas in SXDF308. \label{tab:two}}
\begin{tabular}{lcc}\hline
Parameter		& \multicolumn{2}{c}{Value}\\
\hline
$T_e(O^{2+})$ &\multicolumn{2}{c}{15370 $\pm$ 850 K} \\
$T_e(O^{+})$ (inferred) &\multicolumn{2}{c}{13940 $\pm$  425 K} \\
$T_e(S^{2+})$ (inferred) &\multicolumn{2}{c}{14100 $\pm$ 590 K} \\
$n_e(S^{+})$ &\multicolumn{2}{c}{10$^{+200}_{-10}$ cm$^{-3}$} \\
\hline

\end{tabular}

\end{table}

\end{document}